# Analyzing the Stationarity Process in Software Effort Estimation Datasets

Michael Franklin Bosu[✉]
*Centre for Information Technology*
*Waikato Institute of Technology*
*Hamilton 3240, New Zealand*
michael.bosu@wintec.ac.nz

Stephen G. MacDonell[1] and Peter A. Whigham[2]
*Department of Information Science*
*University of Otago, Dunedin 9054, New Zealand*
[1]stephen.macdonell@otago.ac.nz
[2]peter.whigham@otago.ac.nz

**Abstract**

*Software effort estimation models are typically developed based on an underlying assumption that all data points are equally relevant to the prediction of effort for future projects. The dynamic nature of several aspects of the software engineering process could mean that this assumption does not hold in at least some cases. This study employs three kernel estimator functions to test the stationarity assumption in five software engineering datasets that have been used in the construction of software effort estimation models. The kernel estimators are used in the generation of nonuniform weights which are subsequently employed in weighted linear regression modeling. In each model, older projects are assigned smaller weights while the more recently completed projects are assigned larger weights, to reflect their potentially greater relevance to present or future projects that need to be estimated. Prediction errors are compared to those obtained from uniform models. Our results indicate that, for the datasets that exhibit underlying nonstationary processes, uniform models are more accurate than the nonuniform models; that is, models based on kernel estimator functions are worse than the models where no weighting was applied. In contrast, the accuracies of uniform and nonuniform models for datasets that exhibited stationary processes were essentially equivalent. Our analysis indicates that as the heterogeneity of a dataset increases, the effect of stationarity is overridden. The results of our study also confirm prior findings that the accuracy of effort estimation models is independent of the type of kernel estimator function used in model development.*

## 1. INTRODUCTION

Software engineering datasets emanate from a complex and dynamic ecosystem that involves numerous actions and interactions of people and technologies over time. Observations taken when monitoring projects are used to support decision-making during software development, and the resulting datasets serve as a basis for planning future projects. This paper focuses specifically on software development effort data that may be used in the ongoing management of the cost and/or schedule of current projects as well as in the estimation of the effort required in future projects. Considering the relevance of effort estimates to project schedules and to sound management of the costs incurred by software organizations, it is important to determine any aspects of projects, processes, people, technologies and so on that could have an adverse effect on the accuracy of software effort estimation (SEE). One such aspect is project timing  that is, when in time a project and its constituent activities were undertaken. In ignoring the timing of projects, most current effort estimation practices implicitly assume the underlying development processes to be stationary over time. In regard to SEE, adoption of the stationarity assumption has culminated in the treatment of all past data as equally relevant during the modeling process. The key objective of this paper is to test the validity of this stationarity assumption in the context of SEE.

The range of factors that can affect the effort required in software development is vast. While almost all effort estimation models include parameters that are said to capture the scale and perhaps the complexity of the components to be delivered, there are numerous other influential aspects  the competence and experience of the developers, the participation of the customer, the commitment of top management, requirements ambiguity, adequacy of tools support, communication among the development team, etc. and the list of potential influences is practically endless. The following three studies are just a few of those that have demonstrated the complexity inherent in the management of effort and cost in software projects.

Lagerström et al. [1] studied 31 factors that had the potential to affect the software development cost and productivity of 50 projects in a Swedish bank. They found 10 factors to have a significant influence on the cost and productivity of these projects; moreover, it was noted that not all of these factors are considered in the commonly used effort estimation models. Effort estimation when all the relevant parameters are known is difficult, but it becomes even more challenging when the factors that are needed to derive accurate estimates are not even part of the estimation models being used. Moreover, this result is related to a single organization; the difficulty of transferring knowledge and factors across organizations for software development and effort estimation is additionally complicated.

Wagner and Ruhe [2] divided software productivity factors into two groups; soft factors and technical factors. Soft factors are deemed to be attributes that are influential over the way people work, and include the size of the team, working environment, experience of managers and so on. Technical factors are further divided into three factor subsets: (1) product factors that relate to the software itself, such as the size of the software, the complexity of the software product and the user interface requirements; (2) factors associated with the process used in the development of the software, such as the maturity of the process, how and when prototyping might be used and the management of risk; and (3) development environment factors, such as the programming language used and the tools that support software development. In an earlier study, Maxwell and Forselius [3] assessed the productivity factors of 206 software projects from 26 Finnish companies available in the Experience database. Their analysis [3] identified the company and the type of business of the client organization as being the most influential factors. This finding was contrary to the results of prior studies at the time. On the basis of their findings, they recommended that each organization should keep their own data for project benchmarking purposes; that is, that "nonuniform" data should be used wherever possible.

In light of the above, it will come as no surprise that software effort estimation is itself considered to be a complex and challenging activity. A study by Basten and Sunyaev [4] identified 32 different factors classified into four major groups (estimation process, estimator's characteristics, project to be estimated and external context) as having an influence on the accuracy of software effort estimation. Although the study attributed the most important factors to decision-making by the experts during the estimation process, this approach alone, if not properly managed, could bias the estimation process. It is also very challenging to control up to 32 different variables in order to improve the accuracy of effort estimation.

A potentially important additional aspect missing from the above analyses is that which is in focus here — that is, the stationarity of the development process. It is generally known that software development is a dynamic activity that frequently changes, given its coverage of methodology, technology, team composition strategies and so forth. It is the contention of this study that over some (unknown) period of time an organization's software development processes will not remain static. This is indirectly supported by the fact that since the 1970s there have been proposals for at least five major software development methodologies [6]: the Waterfall Model, Structured Systems Analysis and Design, Formal Methods, the Spiral Model, Rapid/ Joint Application Development and the Agile Methodologies. As a result, a great number of effort estimation techniques have been proposed [7].

Change is thus endemic in software development; it may also be rapid, and yet erratic. In this paper, we therefore assess five software effort estimation datasets to determine whether their underlying processes remain stationary over time.

The rest of the paper is presented as follows. In Sec. 2, we consider the related works. Section 3 describes our research design. Our analysis and results are presented in Sec. 4, and in Sec. 5, we discuss our findings. Threats to the validity of our study are presented in Sec. 6, and in Sec. 7, we conclude the paper.

## 2. RELATED WORKS

This section reviews prior researches that have addressed time-aware SEE models, the use of kernel estimators to assess stationarity in other domains and the application of kernel estimators in stationarity assessment in SEE.

### 2.1. Time-aware SEE modeling

Although numerous SEE models have been proposed (see [7]), the number of studies that have considered project timing information in effort estimation is negligible. This subsection summarizes the few studies that are directly related to the research reported here.

Kitchenham et al. [8] used the start date of projects to split their dataset into four periods prior to building a separate model for each split. The accuracy levels obtained were mixed (as measured by adjusted R-squared). To the best of our knowledge, this was the first study in which project timing was considered in effort estimation. At around the same time, Maxwell [9] used time in two ways in model development — as an explanatory variable and for chronologically splitting the data into training and test sets. Both approaches resulted in good SEE models as measured by the adjusted R-squared. In an analogy-based effort estimation study, Auer et al. [10, 11] used the order of project identifiers in a dataset as a proxy for time in developing models based on a growing portfolio of projects. Although this approach served their purpose by enabling them to detect the volatility associated with project feature dimensions and how it affects the accuracy and reliability of effort estimation across time, it is infeasible to verify whether arrangement of the projects by their IDs represents the chronological order in which the projects were completed (as the ordering of the dataset based on project completion is different from the order of the identifiers for which the data was presented [12]). Five machine learning algorithms were applied by Song et al. [12] to assess the impact of parameter tuning on SEE across different time periods. The results were inconsistent, as some of the algorithms were more sensitive than others to different parameter values. This reinforces the importance of considering the stability of models developed when taking time into account.

A series of studies investigating time-aware SEE have been conducted by Lokan and Mendes [13–15]. Chronological splitting was applied to the ISBSG 10 dataset (http://www.isbsg.org) to assess the time-aware effect of using cross-company data to estimate the effort of single-company projects [13]. Prediction models were assessed against two benchmarks; leave-one-out (LOO) cross-validation and leave two-out cross-validation. Chronological splitting was performed using the start date of the project to be predicted. The authors found no difference in the results of the cross-company and single-company models when evaluated using absolute residuals. (Interestingly, when the authors used z values as their accuracy measures, they found that the cross-company models and the two cross-validation models performed better than the single-company models, reinforcing that conclusion instability can arise due to the use of different accuracy measures when evaluating effort estimation models). Lokan and Mendes [14] then compared further types of chronological splitting in SEE: project-by-project and date-based. The results indicated that neither chronological splitting approach was superior when applied to effort estimation. Using later split dates provided better prediction models than the earlier split dates, leading the researchers to recommend the use of date-based splits in the formation of training and test sets.

The same authors [15] then applied a moving window of the most recently completed projects to new projects in their effort estimation studies. In doing so, they found significant improvement in the accuracy of the moving window models as compared with models built using all available data as the training set. MacDonell and Shepperd [16] assessed the efficacy of two time-aware estimation methods sequential accumulation of projects over time and a constant moving window of size five when applied to a proprietary dataset of 16 projects, and they obtained improved results over project managers' effort estimates, especially for the moving window approach when compared to an LOO model.

Amasaki [17] replicated the use of the previously employed moving window approach [15] on two publicly available datasets, commonly referred to as Kitchenham and Maxwell (see Sec. 3). The window-based approach was found to be superior for both datasets when compared to using all available data in building the effort estimation models, though the estimation accuracy achieved for the Maxwell dataset was better than that obtained for the Kitchenham dataset, implying that the accuracy of the approach may well be dependent on the characteristics of the dataset under consideration.

Two further studies by Lokan and Mendes explored the impact of duration-based moving windows of varying size on the accuracy of effort estimation [18, 19]. The first study used the ISBSG 10 dataset, while the second study replicated the methods employed in the first study when applied to a new dataset (the so-called "Finnish" dataset). For the ISBSG dataset, the accuracy of effort estimation improved for windows of longer duration whilst for the shorter duration windows, the results were not significantly different from the fixed-size windows. Fixed-size windows averagely provided better estimates than duration-based windows for the ISBSG dataset. For the Finnish dataset, neither the duration-based windows nor the fixed-size windows provided any improvement in effort estimation accuracy.

In order to address the nonstationarity associated with software projects, Abrahamsson et al. [5] proposed incremental effort estimation for agile software projects. The incremental model adapts to changes in that a new model is built after the completion of each iteration, which is used to estimate the effort of the next iteration. This incremental approach was applied to two agile projects and compared to the traditional global model. The incremental models were found to exhibit better predictive performance.

Finally, transfer learning was evaluated in the context of analogy-based effort estimation [20] with one of the objectives being to assess the utility of using old data in the models developed for new projects. Two datasets [Constructive Cost Model (COCOMO81) and NASA93] were each split into two to represent different time periods. The results indicated that, if using an instance transfer learner, a substantial number of older observations were transferable over time, leading the authors to advise against the discarding of old data in effort estimation.

### 2.2. GWR and kernel estimator modeling

Fotheringham et al. [21] proposed Geographically Weighted Regression (GWR) as a method to manage autocorrelation and nonstationarity in spatial data. These two properties of spatial data are violated by the assumptions of ordinary least squares (OLS) regression and so hinder the least squares regression from being an effective approach for explaining the relationship between spatial data. GWR derives nonuniform estimates in spatial data; that is, relationships are established in data that belong to a specified (nonuniform) area, as opposed to OLS regression which outputs the estimates of the average or uniform relationships among all observed data. GWR relies on the assumption that entities that are near to each other in a geographical area are more likely to exhibit similar properties than those that are more distant. This assumption is acted on by weighting nearer areas more than distant areas. The nonuniform weighting of records enables the autocorrelation effect across the space to be addressed [22].

Due to its origins in spatial data analysis, GWR has found use as a modeling method in a variety of study contexts. For example, GWR has been used in landscape and urbanization planning studies [23, 24] and in explaining the spatially varying nature of pediatric mortality caused by diarrheal disease in Brazil [25]. Gao and Li [23] used GWR to examine the factors influencing landscape fragmentation and compared the results with those achieved using the OLS regression models, noting that the GWR model results were more accurate in all cases in capturing the nonstationarity of the relationship across space.

The study here employs a procedure similar to GWR wherein nonuniform weightings are applied to software effort estimation data over time. The use of kernel bandwidth values (see Sec. 3.3) also enables the determination of the stationarity of the process underlying the data, except that instead of being applied to the parameters of space, the approach is applied to the parameters of software projects.

### 2.3. Kernel estimator modeling in SEE

In spite of the numerous factors that could dynamically influence the development effort noted above, the rapid pace of change in the field and the proposal of numerous estimation techniques, process (non)stationarity and its effect on SEE have received minimal attention as reported by Smartt and Ferreira [26] in the research literature.

To the best of our knowledge, there are just three prior studies [27–29] in the empirical software engineering domain that have employed kernel estimators in a manner similar to that reported in this paper. Amasaki and Lokan [27] employed a weighted moving windows approach to estimate the effort of software projects using four (Triangular, Epanechnikov, Gaussian and Rectangular) weighted functions. In a subsequent study [28] they applied these same four functions and augmented the approach by applying nonuniform weighting to a growing portfolio of projects. The methods were then compared to nonweighted baseline methods (unweighted moving windows and unweighted growing portfolio). The accuracy of the models was found to be superior for the weighted methods for larger window sizes, while for smaller windows, the results were in favor of the unweighted methods. Though the accuracy of the models is dependent on the size of the windows, in general these results demonstrate the potential utility of weighting projects in software effort estimation. Kocaguneli et al. [29] used five kernel estimators [Uniform, Triangular, Epanechnikov, Gaussian and Inverse Rank Weighted Mean (IRWM)] to estimate software effort based on analogy. The kernel estimators were used to generate nonuniform weights which were applied to 19 SEE datasets. The nonuniform weighted analogies were compared to uniform weighted analogies and the results indicated that uniform weighted analogies provided superior effort estimates. The use of five different kernel estimators did not offer any particular benefit, as all five tended to have similar effects [29].

The study presented here differs from that reported by Kocaguneli et al. [29] in that a wider range of kernel bandwidth values (between 1 and 100) is used in order to discover the stationarity properties of the datasets, whereas five selected kernel bandwidth values were used in the prior study. In addition, this study employs weighted linear regression to build models based on the sequential accumulation of projects according to their completion dates, while Kocaguneli et al. [29] used analogy-based estimation and did not address data accumulation over time. The work presented in this paper has greater similarities with that of Amasaki and Lokan [27, 28] in that it applies linear regression to a growing portfolio of projects using the same set of kernel functions; however, it differs in the use of a wider range of kernel bandwidth values, as they are being applied in this study to assess the stationarity of the datasets, and the processes underpinning the data. The study reported here also employs five datasets exhibiting different characteristics whereas the studies of Amasaki and Lokan [27, 28] used an extract from the ISBSG repository.

Angelis and Stamelos [30] also employed the kernel estimator in software effort estimation based on analogies, however, the purpose of the kernel estimator function is far different from the way it is being employed in this study. They used the kernel function in order to identify the distributions of effort estimates that are not obvious (such as Normal or Log-normal). They used a fixed bandwidth whilst this study uses a range of bandwidths. The following specific research questions are addressed in this study:

RQ1. Do nonuniform models (using nonuniform weighting) result in more accurate software effort estimates than uniform models (using uniform weighting) when applied over time?

RQ2. Does nonstationarity of software engineering datasets affect the accuracy of effort estimation models when applied over time?

RQ3. What is the effect of different bandwidths (if any) on software effort estimation model accuracy?

RQ4. Does kernel type affect the accuracy of software effort estimation models?

## 3. RESEARCH DESIGN

In this section, we first describe each of the five datasets to be analyzed along with the particular computation of effort estimation used in each case. Four of the datasets have been publicly available and so have been used extensively in prior researches. We then describe our model development and evaluation process before specifying how the various kernel weightings are determined.

### 3.1. Dataset descriptions

#### 3.1.1. NASA93 dataset

The NASA93 dataset was collected by NASA from five of its development centers and it collectively represents 14 different application types. The entire dataset comprises 93 projects (hence its name) undertaken between 1971 and 1987. According to Lum et al., [31] projects were completed in the years indicated in the versions of the dataset that are available from the PROMISE Repository, http:// openscience.us/repo/.

There is no information in the dataset as to when projects were started; however, the completion dates alone provide some of the information needed in order to build models of effort for subsequent years' projects. Even when projects have already started, a model based on completed projects from the year before, or prior, could be used to update early or current estimates.

Table 1. Definitions of COCOMO81 effort multipliers.

| Increase these to decrease effort | acap: analyst capability |
| --- | --- |
|  | pcap: programmer capability |
|  | aexp: application experience |
|  | modp: modern programming practices |
|  | tool: use of software tools |
|  | vexp: virtual machine experience |
|  | lexp: programming language experience |
|  | sced: required development schedule |
| Decrease these to decrease effort | data: database size |
|  | turn: computer turnaround time |
|  | virt: virtual machine volatility |
|  | stor: main memory constraint |
|  | time: execution time constraint |
|  | rely: required software reliability |
|  | cplx: product complexity |

*Source*: See [32].

The dataset is structured according to the COCOMO81 format developed by Boehm [32]. It comprises 24 attributes of which 15 are the mandatory effort multipliers. Tables 1 and 2 describe the effort multipliers and project development modes of COCOMO81, respectively. Effort multipliers are assigned a range of predefined values which were obtained from regression analysis of the original COCOMO81 data.

The other attributes of relevance are product size, measured in thousands of lines of code (KLOC), and effort, measured in calendar months (where one calendar month is said to be equivalent to 152 person-hours of effort). The computation of effort for COCOMO81 projects is given by

$$effort(personmonths) = a * (KLOC^b) * \left(\prod_i EM_i\right), \quad (1)$$

where KLOC is the size measured in thousands of lines of code and EM represents the effort multipliers. COCOMO81 projects are classified into three development modes such that each requires the use of certain parameter values in the model: the values of a and b are domain-specific ones dependent on the mode of the project being developed, as represented in Table 2.

### 3.1.2. Desharnais dataset

The Desharnais dataset was collected from 10 organizations in Canada by Jean-Marc Desharnais. This dataset is made up of 81 projects conducted between 1983 and 1988. The dataset is described by 12 attributes whose size and effort attributes were measured using function points and person-hours, respectively. In most studies that employ this dataset, 77 of the 81 records are used because of missing data in four records [34]. In this study, the version with 77 projects is therefore also used. Similar to the NASA93 dataset, only the year of project completion was recorded for the Desharnais dataset, therefore, the training and test datasets are formed in the same way as the NASA93 dataset (i.e. by using the year of project completion).

Though there are 12 attributes in the Desharnais dataset, analysis carried out by Desharnais himself [35] identified the size and language attributes as those that are influential in a regression model. Kitchenham and Mendes [36] supported Desharnais' claim by proposing the use of the language attribute as a dummy variable. This approach has therefore been adopted in this study for the models developed for this dataset, as shown in the following equation:

$$ln(effort) = ln(size) + language \quad (2)$$

This study used the adjusted function points value as the most complete-size attribute (rather than the raw function point count). The language attribute in Eq. (2) is a dummy variable made up of three distinct values for which the Basic Cobol project is represented as "1" in the Desharnais dataset. This (Basic Cobol project) has been chosen as the reference dummy variable in this study.

Table 2. Standard COCOMO81 development modes.

| Mode | a | b | Explanation |
|---|---|---|---|
| Organic | 3.2 | 1.05 | Projects from relatively small software teams developing software in a highly familiar, in-house environment. |
| Embedded | 2.8 | 1.20 | Projects operating within (embedded in) a strongly coupled complex of hardware, software, regulations and operational procedures. |
| Semi-detached | 3 | 1.12 | An intermediary mode between organic and embedded. |

Source: See [33].

### 3.1.3. Kitchenham dataset

The Kitchenham dataset [8] was collected from the American-based multinational company Computer Sciences Corporation (CSC). This dataset contains information about 145 software development and maintenance projects that CSC undertook for several clients. There are 10 attributes considered, the size attribute was measured in function points, and effort was measured in person-hours. The attributes also include the start date and estimated completion dates, and the projects were undertaken between 1994 and 1999. The attributes useful for effort modeling (based on prior research evidence) are the size attribute and the application type attribute. This study used the application type attribute as a dummy variable with the reference value being the "Development" type. Again following prior work, this study uses 105 records related to projects developed for the so-called "client 2" [17].

As this dataset includes information about the actual start date of projects and their duration in days, these values are used to compute each project's completion date. Training sets are formed based on the years in which projects were completed, as was done for the NASA93 and Desharnais datasets. Composition of the test datasets follows a slightly different process, however, because of the availability of actual start dates: a test set consists of projects completed in the subsequent year and started after the date the last project in the training set was completed. This dataset consists of 67 perfective maintenance projects and 38 development projects. The model formulation is shown as follows:

$$ln(effort) = ln(size) + type. \quad (3)$$

### 3.1.4. Maxwell dataset

The Maxwell dataset, collected from a Finnish commercial bank, is made up of 62 projects with 27 attributes [9]. The size attribute was measured in function points. This dataset enables the estimation of project effort measured in person-hours. The projects were undertaken between 1985 and 1994. The Maxwell dataset has both the actual start and completion dates and, as such, the training and test sets were formed in the same manner as described for the Kitchenham dataset. Prior studies [9, 17] have found the explanatory variables requirement volatility (T08) and quality of requirements (T09) in addition to the size attribute to be of value in effort estimation. This study therefore adopts these variables in model development as shown in the following equation:

$$ln(effort) = ln(size) + T08 + T09 \quad (4)$$

### 3.1.5. XBC dataset

The proprietary XBC dataset characterizes the development of 16 systems built using the C and C++ programming languages, for deployment in either a runtime environment or a Unix-based nonruntime environment. The projects followed a waterfall approach with data recorded as labor hours for 10 different process phases; Project planning, Requirement specification, Training/Learning, Design specification and documentation, Implementation, Test and test specification, Release, Installation & Manuals, Maintenance and Process assurance. Records were kept for the time spent on project management, and the total labor hours for each project were also noted. In this dataset, two estimates of the managers were also stored against the above-mentioned phases for each project: an original estimate, which refers to the initial estimate of the project manager, and a current estimate, a revised estimate which at any point in time is the most up-to-date estimate. Counts of the number of requirements and the commented lines of code (CLOC) were recorded to provide (early and later) indicators of the size of the projects. Other project information recorded included a project identifier and a name for each project, as well as the names of the project managers involved. Project completion dates and their durations, in weeks, were available. Models built for this dataset followed the same procedure as that used for the NASA93 and Desharnais datasets in terms of employing the project completion dates. All 16 projects were undertaken within a 22month period, between October 1999 and August 2001. Preliminary analysis identified the managers' original effort estimate(org effort) and the number of requirements as potential explanatory variables in a linear regression model to estimate the total actual effort (total effort). Subsequent correlation analysis led to the use of only the managers' original estimate, with a correlation coefficient of 0.83 with the total effort. The following equation is therefore used to estimate the effort for this dataset:

$$ln(total\_effort) = ln(org\_effort). \quad (5)$$

### 3.2. Effort estimation model development

In software effort estimation modeling, as in many other fields, the (secondary) dataset is usually split into two, forming a larger training set and a smaller test set. Models are then built using the training set, and the unbiased performance of the models is evaluated on the test set. This study follows a similar approach; the specifics of how the training and test sets are formed are described in Sec. 3.2.1. All models in this study are developed using the statistical package R (version 3.0.2). In preparatory testing, the Shapiro–Wilk test of normality was applied to the numeric variables in the training sets. All such variables that failed the normality test were logarithm transformed, meaning that in the associated models developed, log(effort) [shown as ln(effort) in the equations] would be the dependent variable and log(size) [ln(size)] one of the explanatory variables. The estimated (natural log) effort values are back-transformed to unscaled values prior to the computation of any accuracy measures. All models are developed using linear regression, considered to be a baseline modeling approach in effort estimation [7]. The actual linear regression equations for each dataset have been presented in Sec. 3.1. Models constructed in this study are all well-formed models which imply that the degrees of freedom are considered in the formation of the training data, whereby, models are built only when the number of projects is at a minimum, two plus the number of explanatory variables.

#### 3.2.1. Modeling algorithm

This paper generally follows the sequential accumulation approach used by MacDonell and Shepperd [16] in forming the training sets for the effort estimation models. As such, the following procedures are applied to all datasets modeled in this study:

**Step 1.** For each dataset with timing information, select the first year/month in which projects were completed as the training set — if the first year/month of projects comprises fewer than the number of observations needed to build a well-formed model, add the next year(s)/month(s) of projects until the minimum requirement for a well-formed model is satisfied. The subsequent year/month of projects is used as the test set.

**Step 2.** Check for normality in the distributions of the training data — if the data follow a normal distribution go to step 3, else go to step 2.1.

**Step 2.1.** Apply the appropriate transformation to make the data normal and recheck normality for verification as above.

**Step 3.** Build a regression model using the training data.

**Step 4.** Apply the model obtained in step 3 to predict the values in the test set.

**Step 5.** Calculate the accuracy measures (see below) for the prediction model.

**Step 6.** Add the test year's/month's data to the training set, and the subsequent year's/month's data becomes the new test set.

**Step 7.** Repeat steps 2–6 until the estimation of the last year/month of projects.

#### 3.2.2. Model evaluation

We employ the relative error (RE) measure in evaluating each of the models developed in this study. This is because the relative error measure accounts for the variability in data and as such it is robust to outlier data points. Values of RE equal to or greater than 1 indicate that the model is performing no better than the prediction of constant value [37] while values approaching zero indicate an increasingly accurate prediction. The relative error is computed using the following equation:

$$RE = variance(residuals)/variance(test\ data). \quad (6)$$

In order to apply a consistent approach to our analysis, the completion date of each project in the five datasets is the only property of time considered in the determination of the kernel weights in this subsection [even though three of the datasets (Kitchenham, Maxwell and XBC) include project start and completion dates].

Table 3 shows the four kernel estimators used in generating weights applied to the datasets (where the Uniform kernel serves as a nonweighted baseline). To find $t$, we used

$$t_{ij} = (t_j - t_i)/b, \quad (7)$$

Table 3. Formulae of kernel types.

| Kernel type | Formula |
|---|---|
| Uniform | $W_{ij} = 1, |t| < 1$ |
| Gaussian | $W_{ij} = \exp(-0.5 \times t^2), |t| < 1$ |
| Epanechnikov | $W_{ij} = 1 - t^2, |t| < 1$ |
| Triangular | $W_{ij} = 1 - |t|, |t| < 1$ |

where $t_{ij}$ is the period, or in this case, the number of $years/months$ that have elapsed between project $i$ and the target project $j$ (i.e. the project being estimated). $W_{ij}$ is the weight applied to project(s) completed in year/month $i$ with reference to the projects in a target $year/month\ j$, and $b$ is the kernel bandwidth (discussed later in this subsection).

The value of 1 is assigned to the oldest completion period in each dataset (note that the completion period is month for the XBC dataset, whilst it is year for the four other datasets) and a yearly increment of 1 is applied thereafter. The elapsed time periods are determined between a specific year and the target year to be used in the application of the formulae in Table 3 to derive the weights for projects in specific years; each past year is subtracted from the target year and the results indicate the elapsed time (in years or in months for the XBC dataset) from the target year. For instance, given two projects developed in different years, $t_{ij} = j - i$. The weight is 1 when $i$ is equal to $j$. This approach to weight generation applies to four of the datasets (NASA93, Kitchenham, Maxwell and Desharnais) that represent the completion year parameter with integer values. In the XBC dataset, the same procedure as previously discussed is used, the only difference being that the month of project completion is used, since all 16 projects were completed within a short period of 18 months. The value for the month parameter starts with 0.1 for the oldest project in the set and a monthly increment of 0.1 is made thereafter. Projects completed in the same month are assigned the same value, just like projects that were completed in the same year in the other four datasets. The bandwidth controls the weighting contribution of neighboring projects, i.e. projects from specific years or months. The selection of the optimum bandwidth is important as it is known to be more influential than the choice of the kernel function [24] as it affects the extent of decay of the kernel function. Previous studies [38, 39] conducted comprehensive reviews of the various approaches that can be used to determine the optimum bandwidth for kernel estimator functions. Whilst Jones et al. [38] identified the "solve the equation plug-in" method as the best kernel bandwidth value selection approach, Turlach [39] could not settle on any particular method, but rather presented the strengths and weaknesses of each method under various conditions. It is not the intent of this paper to address the "best" bandwidth selection problem here, as our core interest is in the stationarity of the software engineering process over time.

Figure 1 depicts the weights that are generated for selected bandwidth values for the datasets based on the Gaussian kernel used in this paper (note that for clarity, it is impractical to show all the bandwidth values between 1 and 100). Figure 1(e) differs markedly from the other graphs, illustrating the more compact nature of the completion periods between the XBC projects, as the weights over time for all the different bandwidth values are close compared to those for the other four datasets. For this study, the bandwidths are set between 1 and 100 at increments of 1. Figure 1 shows that as the bandwidth value increases, the weights applied to all projects in the training set approach 1. Older projects have smaller weights because the assumption is that the underlying software process used in generating the data is different from that used for current projects. It is also evident in Fig. 1 that small bandwidth values such as 1 and 2 lead to a rapid decline in the weights

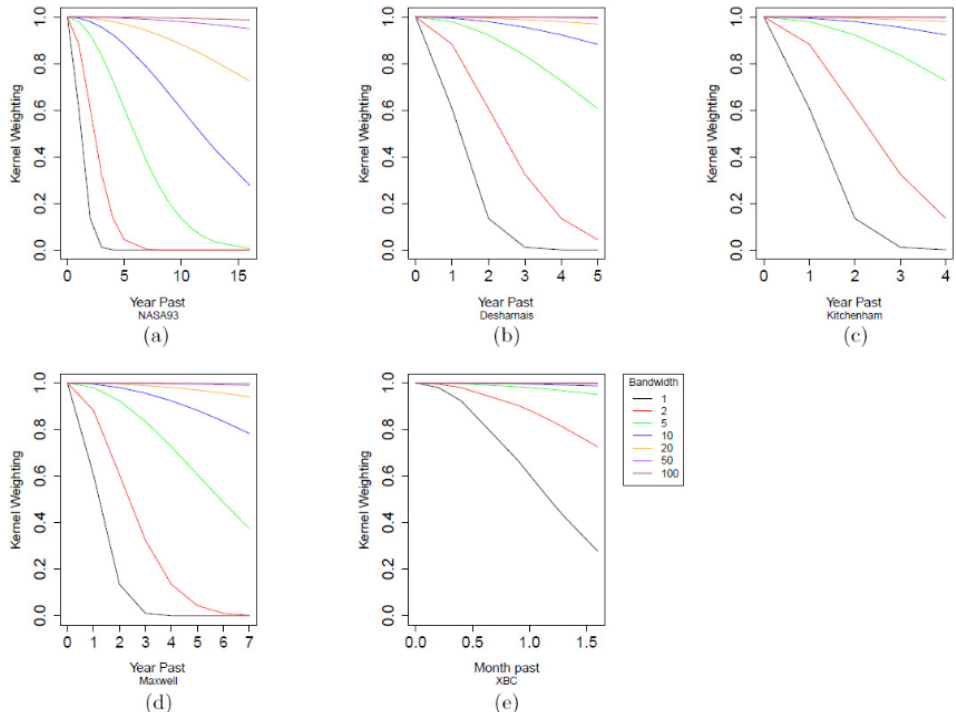

Fig. 1. Weights generated for datasets using the Gaussian kernel.

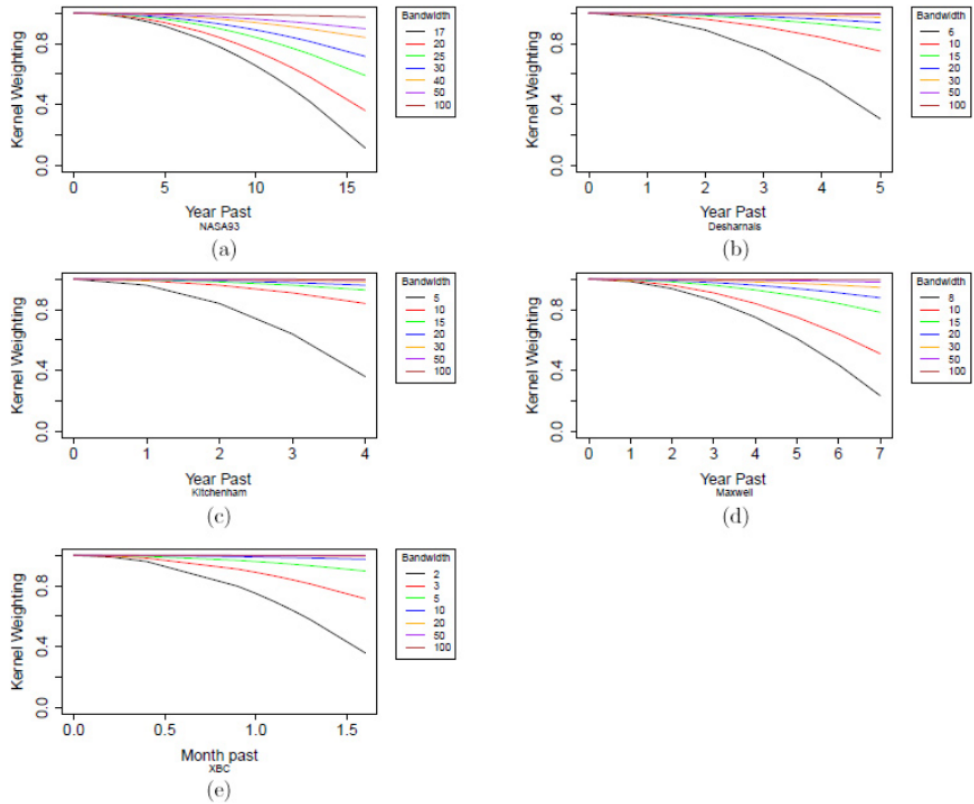
Fig. 2. Weights generated for datasets using the Epanechnikov kernel.

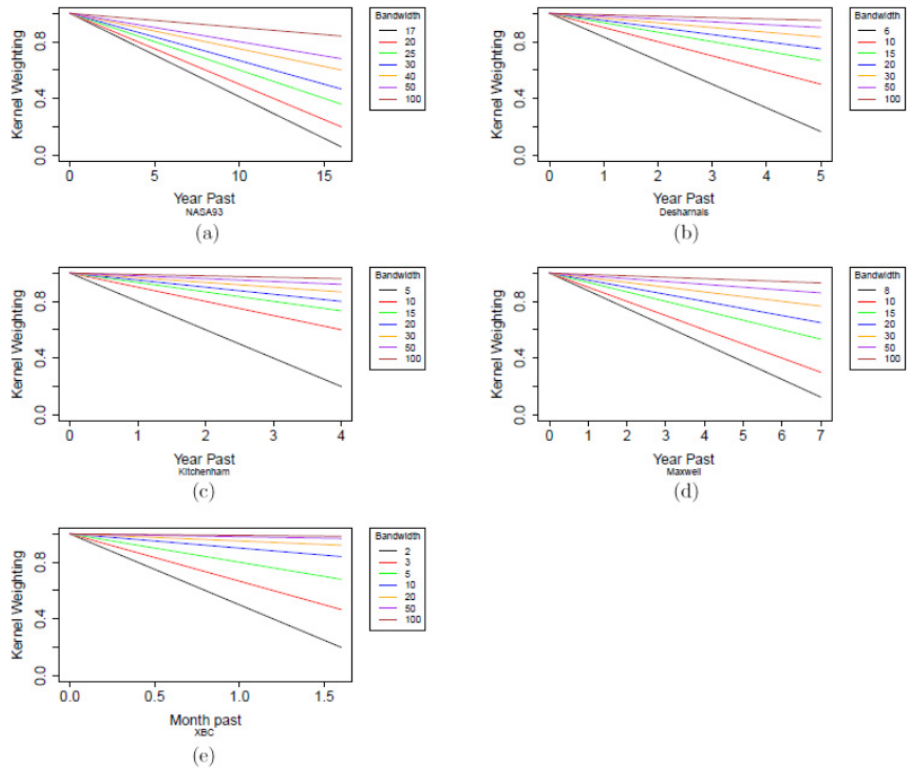
Fig. 3. Weights generated for datasets using the Triangular kernel.

that are assigned to projects that occur later in time from the target year. However, the weight for larger bandwidth values declines gradually and as such, the weights for the data in the training set become nearly the same irrespective of the completion date of a project.

Figure 2 shows the weights generated for selected bandwidth values for the datasets based on the Epanechnikov kernel. The concave nature of the curves corresponds to the expected shape of this particular kernel [28].

In comparison to the Gaussian kernel curves, the weights decrease a little more gradually, for all bandwidth values across the periods of project completion. Finally, Fig. 3 shows the weights generated for selected bandwidth values for the datasets based on the Triangular kernel. The graphs are linear for all bandwidth values and across all periods. Just like the Gaussian and the Epanechnikov kernels, the weights for larger bandwidth values decline in a more gradual manner.

## 4. ANALYSIS AND RESULTS

The kernel weights generated as per the procedure described in Sec. 3.3 are applied to the effort estimation models for the five datasets. The relative errors of the models are computed over the specified range of bandwidth values as shown in the graphs reported in this subsection. The use of the kernel functions enables the application of nonuniform weights to the projects in these datasets as they are used to develop effort estimation models. In order to determine the stationarity or otherwise of these datasets, effort estimation models are developed according to the modeling algorithm of Sec. 3.2. The modeling equations derived for each of the datasets in Sec. 3.1 are subsequently applied.

In order to determine whether a model exhibits a stationary process, the weight graphs in Sec. 3.3 should be considered alongside the graphs depicting prediction errors that follow. For example, in the case of the Gaussian kernel, Fig. 1 is read in combination with graphs of the models developed for each of the five datasets that used the Gaussian kernel in weight generation, shown in Fig. 4. The bandwidth at which stationarity was attained is identified on the graph of the respective dataset and then this bandwidth value is mapped onto the corresponding curve in Fig. 1 to determine the year or period at which the models remained stationary. This process is repeated for the Epanechnikov kernel and the Triangular kernel in the interpretation of the results.

In the plots presented in this section, the accuracy measure of the model built using the weights generated by the kernel estimators is shown on the plots as "train", which is effectively the nonuniform model (applying nonuniform weighting). The nonuniform model is then used to predict the effort of projects in the test set, indicated as "test" on the graphs. Similarly, the result of the uniform model (where no weighting is applied) is indicated on the plot as "train global", and the model is then used to predict the effort of projects in the test set, indicated as "test global". The results are shown on each graph to aid comparison of the models and to enable the identification of models that are stationary or otherwise. It is worth noting that, in presenting the results, emphasis is placed on the training model outcomes because the intention is to identify the stationarity status in the data. The results are subsequently presented for each of the datasets in this section.

### 4.1. NASA93 dataset
The results of the models developed for the Gaussian kernel modeling of the NASA93 dataset are shown in Fig. 4. The graphs show the relative errors against bandwidth values for models built over the various time periods under consideration.

In Fig. 4(a), at approximately a bandwidth of 5, the nonuniform model and the uniform model converge, meaning a stationary process is achieved at this point. Looking at a bandwidth of 5 on Fig. 1(a) indicates that convergence would occur at about the 15th year of projects in the training set. Given that the training set for this model is made up of only seven years of projects, this means there is effectively no convergence, implying that these projects exhibit a nonuniform process. The underlying process can therefore be said to be nonstationary. The predictions for the 1980 projects based on both the nonuniform model and the uniform model are large (greater than 1) which could be due to a number of reasons; such as the training data being different from the data or the models themselves being not good.

The inclusion of the 1980 projects in the model of Fig. 4(b) led to the convergence of the nonuniform model to the uniform model starting at a bandwidth value of 3 and the actual convergence occurring at a bandwidth of 8. Since according to Fig. 4(a) bandwidth values of 5 or greater converge beyond the number of years of projects in the entire NASA93 dataset, this indicates that the model shown in Fig. 4(b) also exhibits a nonstationary process, because the training data only consists of eight years of projects. The models' predictions, however, were improved when compared to Fig. 4(a).

The results of the model depicted in Fig. 4(c) are similar to those shown in Fig. 4(a). These two models, Figs. 4(a) and 4(c), converge at about a bandwidth value of 5. According to Fig. 1(a), a bandwidth value of 5 converges beyond the number of years that constitute the entire NASA93 dataset, implying that the model of Fig. 4(c) also exhibits a nonstationary process. The prediction of the 1983 projects based on the nonuniform model also produced relatively large errors just as was observed for the model of Fig. 4(a). Figure 4(d) indicates that at about a bandwidth of 14, the model started converging and that the actual convergence occurred at a bandwidth of 25, which according to Fig. 1(a) is well beyond the number of years of projects that constitute the training set, implying that all of the projects that constitute the nonuniform model exhibited a nonstationary process. However, the predictions based on both the nonuniform model and the uniform model are good with RE values less than 1 as indicated on Fig. 4(d).

The nonuniform model of Fig. 4(e) started approaching a stationary process at a bandwidth value of about 17. If this is mapped onto Fig. 1(a), it is beyond the number of years for which convergence can be attained based on the training set, implying that the model exhibits nonstationary characteristics. For these projects also, the predictions based on both the nonuniform model and the uniform model are good (RE values less than 1).

The nonuniform models of Figs. 4(f) and 4(g) both started approaching the curve of the uniform model at a bandwidth value of around 20. The actual convergence of the nonuniform models to the uniform models occurred at the bandwidths of 30 and 35, respectively, in Figs. 4(f) and 4(g). This again occurs beyond the number of years of projects in the datasets [as indicated in Fig. 1(a)], which implies that the projects used in building the models exhibited nonstationary characteristics. For the model shown in Fig. 4(f), the predictions based on both the nonuniform model and the uniform model are good with RE values less than 1. In Fig. 4(g), the predictions based on both the nonuniform model and uniform model are good with an RE of less than 0.2.

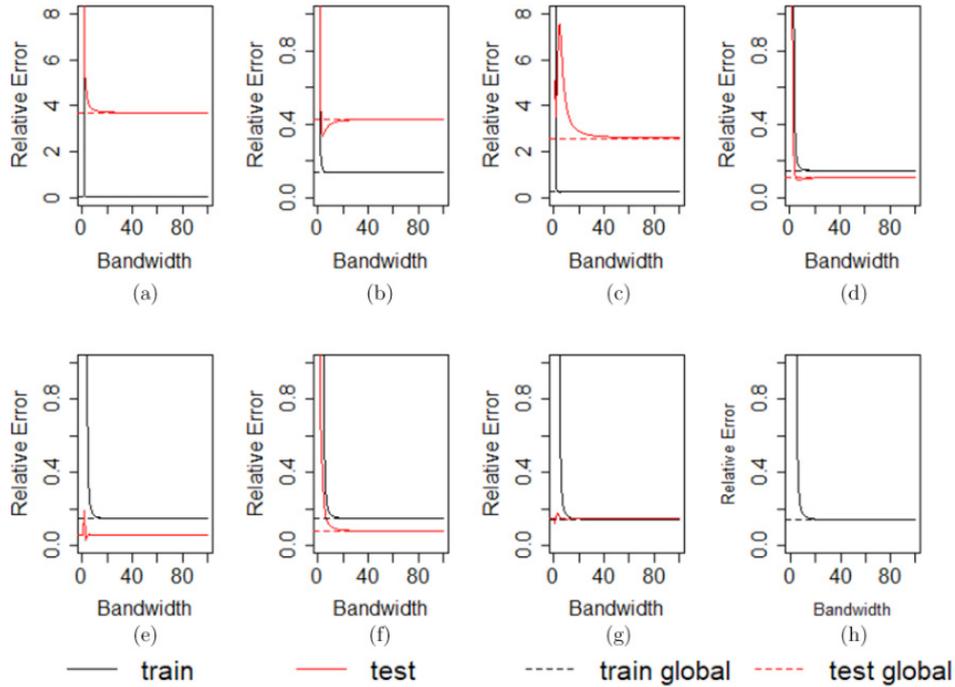

Fig. 4. Gaussian models: Relative error against bandwidth for the NASA93 dataset. (a) All projects up to 1979 as the training set and 1980 as the test set. (b) All projects up to 1980 as the training set and 1982 as the test set. (c) All projects up to 1982 as the training set and 1983 as the test set. (d) All projects up to 1983 as the training set and 1984 as the test set. (e) All projects up to 1984 as the training set and 1985 as the test set. (f) All projects up to 1985 as the training set and 1986 as the test set. (g) All projects up to 1986 as the training set and 1987 as the test set. (h) Model built with all NASA93 data.

A model developed using the entire NASA93 dataset, as shown in Fig. 4(h), started approaching the uniform model curve at a bandwidth of 15 and actually converged to that of the uniform model at about a bandwidth of 18. This convergence point according to Fig. 1(a) requires more than 14 years of projects that constitute the NASA93 dataset, implying that the process underlying this model is nonstationary.

Overall, Fig. 4 indicates that the accuracy of the uniform models is better than (i.e. they exhibit lower relative error values) the nonuniform models for the NASA93 dataset. The curves also show the existence of nonstationary processes underlying the projects of the NASA93 dataset across the different projects over time, evident from the rapid decline of the relative error of the nonuniform models as the bandwidth value increases.

Figure 5 depicts the results of the models built using the Epanechnikov kernel for the NASA93 dataset. The bandwidth values for the Epanechnikov kernel and the Triangular kernel models for this dataset were set between 17 and 100 as lower values are not possible as per the formulae. Figures 5(a)–5(d) exhibit a stationary process, in line with their equivalent Gaussian models using the same bandwidth values. Figures 5(e)–5(h) exhibit some initial level of nonstationarity which is similar to the equivalent Gaussian models (i.e. using bandwidth values between 17 and 100) as these models attained convergence to stationary status at higher bandwidth values that have been explained previously for Figs. 4(e)–4(h) in this subsection. The results of the models and their predictions are similar to those obtained for the Gaussian models shown in Fig. 4.

In the interests of parsimony, the graphs of the Triangular models for the NASA93 dataset are not shown here as the results achieved are similar to those obtained for the Epanechnikov models[1].

The accuracy of the nonuniform models is not obviously different across all time periods for the NASA93 dataset, as shown in Figs. 4 and 5. This implies that the addition of new projects data to build subsequent models is neither having positive nor negative effects on the accuracy of the models (as relative errors are nearly the same for all the models of the NASA93 dataset) across time. It is worth noting, however, that the prediction performance of the models is different across time as some of the models are estimating better than others as indicated in Figs. 4 and 5. These same effects have been observed for the uniform models for this dataset.

Although a few of the models based on the Epanechnikov kernel and the Triangular kernel appear to exhibit an underlying stationary process (due to the kernel bandwidth values starting at 17 instead of 1 in the case of the Gaussian kernel), the predominantly nonstationary process for the NASA93 dataset can be attributed to the fact that the projects, though conducted under the auspices of one organization (NASA), were developed by several external contractors with potential practice diversity. The applications themselves were also very diverse, comprising 14 application types and spanning a development period of 16 years.

---

[1] Note, however, that all graphs that have not been shown in this paper are available online.https:// tinyurl.com/IJSEKE2020-Stationary-Analysis.

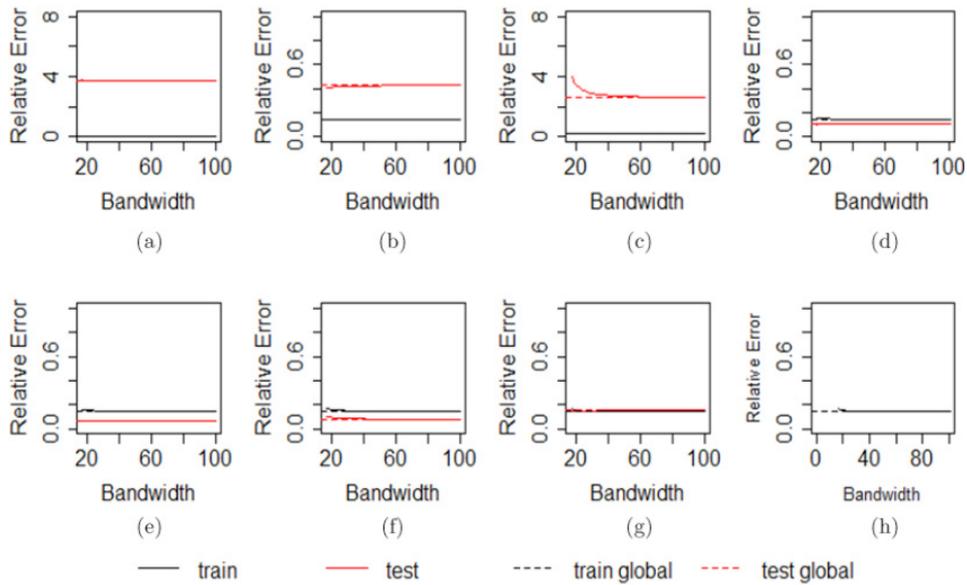

Fig. 5. Epanechnikov models: Relative error against bandwidth for the NASA93 dataset. (a) All projects up to 1979 as the training set and 1980 as the test set. (b) All projects up to 1980 as the training set and 1982 as the test set. (c) All projects up to 1982 as the training set and 1983 as the test set. (d) All projects up to 1983 as the training set and 1984 as the test set. (e) All projects up to 1984 as the training set and 1985 as the test set. (f) All projects up to 1985 as the training set and 1986 as the test set. (g) All projects up to 1986 as the training set and 1987 as the test set. (h) Model built with all NASA93 data.

### 4.2. Desharnais dataset

Figure 6 indicates the relative error values across different time periods and band-widths for the Desharnais dataset using the Gaussian kernel function. These results indicate that, in general, the uniform models are nearly the same as the nonuniform models in terms of their accuracy, though the nonuniform models are marginally better in a few cases as shown in Figs. 6(b) and 6(c). The predictions based on both the uniform and nonuniform models for this dataset are good in terms of accuracy as they all recorded an RE of less than 1. In Fig. 6(e), which shows the model built with the entire Desharnais dataset, the nonuniform and the uniform model results are nearly the same, with both exhibiting an underlying stationary process. Taken overall, the results of the Desharnais model analysis generally indicate a nearly stationary process across different bandwidths and across time. For this dataset, the nonuniform model and uniform model predictions are nearly the same, for all the models.

The results obtained from the models developed using the Epanechnikov kernel and the Triangular kernel are not shown here because they are largely similar to those obtained for the Gaussian models.

Across time, the accuracy of both the nonuniform and uniform models improved in the case of Fig. 6(b) when new projects were included in building the model [this being an improvement upon the model of Fig. 6(a)]. The accuracies of the models of Figs. 6(b)–6(e) remained the same, however. This implies that the addition of new projects in building subsequent models as shown in Figs. 6(b)–6(e) had no impact in terms of model accuracy. The predictions based on the models (nonuniform and uniform) are similar to those described for the NASA93 dataset in Sec. 4.1, as some of the models' predictions are better in terms of accuracy than the others across time.

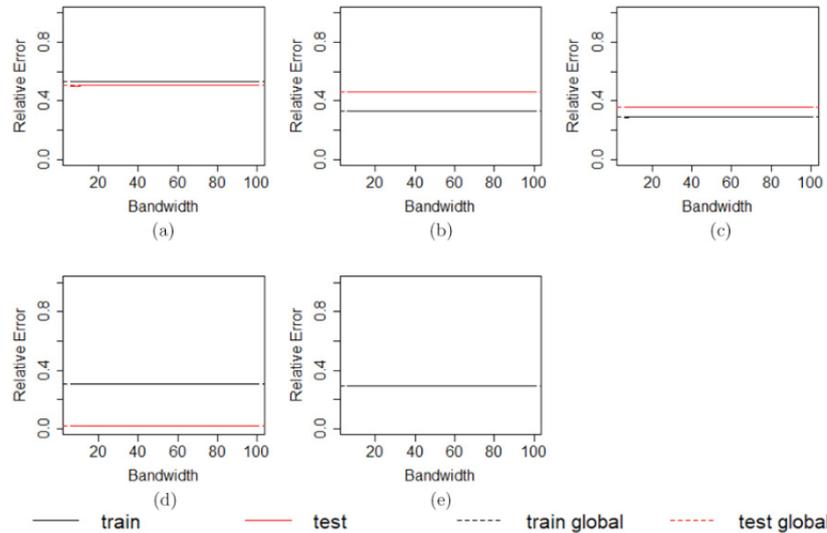

Fig. 6. Gaussian models: Relative error against bandwidth for the Desharnais dataset. (a) All projects up to 1984 as the training set and 1985 as the test set. (b) All projects up to 1985 as the training set and 1986 as the test set. (c) All projects up to 1986 as the training set and 1987 as the test set. (d) All projects up to 1987 as the training set and 1988 as the test set. (e) Model built with all Desharnais data.

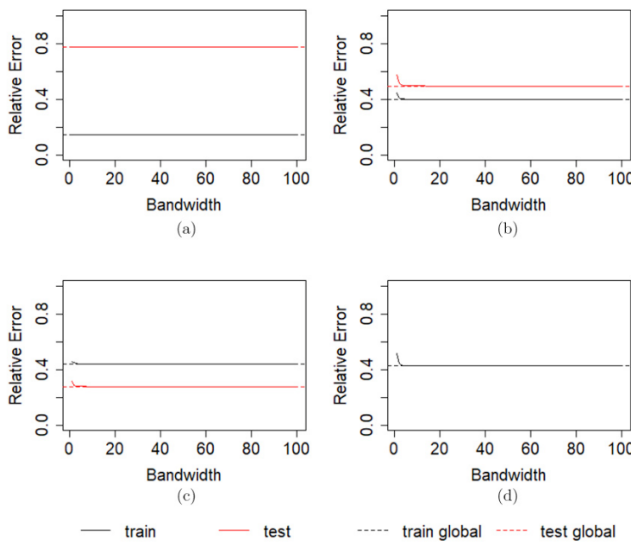

Fig. 7. Gaussian models: Relative error against bandwidth for the Kitchenham dataset. (a) All projects up to 1995 as the training set and 1996 as the test set. (b) All projects up to 1996 as the training set and 1997 as the test set. (c) All projects up to 1997 as the training set and 1998 as the test set. (d) Model built with all Kitchenham data.

The relative stationarity of the models built for the Desharnais dataset is somewhat surprising because this dataset was collected from 10 different organizations in Canada over a period of seven years. The homogeneity of these models may be due to the fact there were few project types and programming languages used in the development of the projects in the Desharnais dataset. In addition, it is also possible that different organizations working during the same period may use similar practices which contributed to the stationarity models exhibited by the Desharnais dataset.

**4.3. Kitchenham dataset**

Figure 7 depicts the graphs for the models developed using the Gaussian kernel when applied to the Kitchenham dataset. Figure 7(a) depicts a near stationary process. Figures 7(b)–7(d) exhibit nonstationarity at lower bandwidth values until they converge to a stationary process at the bandwidth values between 4 and 10. Mapping these bandwidth values onto Fig. 1(c) indicates that the models will not attain stationarity in time thus, the processes underlying the models developed for Figs. 7(b)–7(d) are interpreted as being nonstationary. The results for this dataset are therefore mixed there is evidence of a stationary process in one of the models while the other three imply a nonstationary process. The predictions based on the Gaussian model are relatively good for this dataset as they all attained a relative error of less than 1.

The bandwidth values for the Epanechnikov kernel and the Triangular kernel models for this dataset were set between 5 and 100. The Gaussian models using these same bandwidth values exhibited near stationary processes for all the models. The results of the models based on the Epanechnikov and Triangular kernels are not shown here, they are similar to their equivalent Gaussian curves based on the exact bandwidth value comparisons (they exhibit similar stationarity and nonstationarity at the same bandwidth values, respectively).

The accuracies of the models of all three kernel estimators are similar for the Kitchenham dataset as depicted by their respective curves. Both the respective nonuniform models and the uniform models generated similar results in terms of the RE measure.

Model accuracy (for both the nonuniform and uniform models) worsened upon the introduction of new projects to build the model of Fig. 7(b) [i.e. the model based on Fig. 7(b) is worse than the model of Fig. 7(a)]. The models further worsened across time [i.e. from Fig. 7(b) to Fig. 7(c)] and remained the same, i.e. Fig. 7(c) is the same as Fig. 7(d), in terms of relative error. The predictions based on the models differ across time as was observed for the previous two datasets.

The mixed results (both stationary and nonstationary models) obtained for the Kitchenham dataset could be attributed to the different practices associated with the development types: it seems likely that the organization would have applied different processes to new software development projects as compared to its perfective maintenance projects. This could explain the nonstationarity of some of the models. On the other hand, the stationary model could be due to the fact that all projects were developed by one organization for a single client, and as such, similar general (organization-level) procedures could have been applied.

**4.4. Maxwell dataset**

Figure 8 illustrates the models built using the Gaussian kernel for the Maxwell dataset. Figures 8(a) and 8(b) are near-stationary. Figures 8(c) and 8(d) attained stationarity at about the bandwidth of 5 and 12, respectively.

Mapping these bandwidth values to Fig. 1(d) indicates that no stationary process will be attained, thus these two models exhibit nonstationary processes. For the Maxwell dataset, it is therefore evident that while two models are stationary, the other two are nonstationary, indicating that in general the Maxwell dataset reflects a diversity of projects and practices over time.

The results of the models obtained based on the Epanechnikov kernel (Fig. 9) and the Triangular kernel (Fig. 10) (using the bandwidth values between 8 and 100) are similar to those achieved with the Gaussian kernel as indicated in Fig. 8; at a bandwidth value of 12, all the Gaussian curves of Fig. 8 have converged into a stationary process. Overall, the models' results according to relative error were satisfactory, as all the models resulted in a relative error value of less than 1. The predictions based on both the nonuniform models and the uniform models are almost identical for all the models developed for the

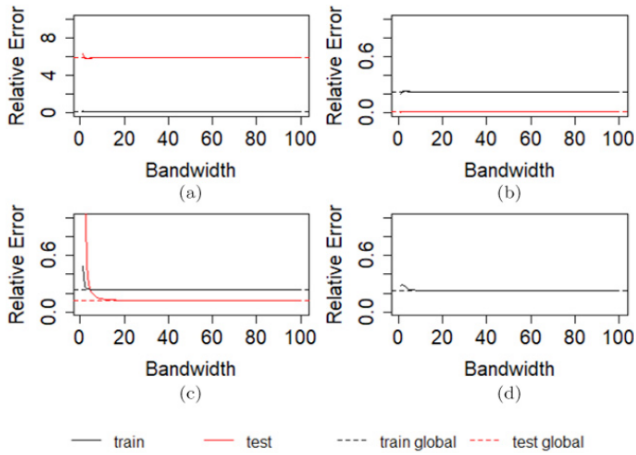

Fig. 8. Gaussian models: Relative error against bandwidth for the Maxwell dataset. (a) All projects up to 1988 as the training set and 1989 as the test set. (b) All projects up to 1989 as the training set and 1990 as the test set. (c) All projects up to 1990 as the training set and 1991 as the test set. (d) Model built with all Maxwell data.

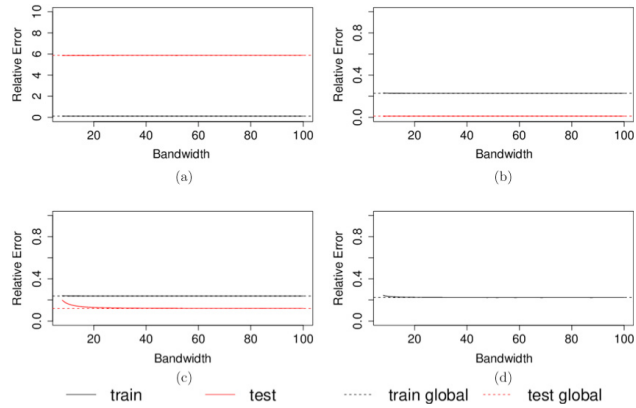

Fig. 9. Epanechnikov models: Relative error against bandwidth for the Maxwell dataset. (a) All projects up to 1988 as the training set and 1989 as the test set. (b) All projects up to 1989 as the training set and 1990 as the test set. (c) All projects up to 1990 as the training set and 1991 as the test set. (d) Model built with all Maxwell data.

Maxwell dataset as shown by the respective kernel type curve. These predictions are relatively good with a relative error of less than 1 for all the models with the exception of the first model built for each kernel type.

The model accuracy (for both nonuniform and uniform models) is the same upon the introduction of new projects to build subsequent models (especially for the uniform models and after the convergence of the nonstationary models) across time as evidenced in Figs. 8–10. The predictions based on the models differ across time as was observed for the previous three datasets.

The stationarity underlying some of the models of the Maxwell dataset can be attributed to the fact that the projects were developed internally in a single organization, and as such, some of the processes might be similar across projects and time. The presence of some nonstationary models, in contrast, could be due to the fact that these projects consisted of four different application types, with the consequent potential of different processes and methodologies being used. Some projects also used code generators whilst others did not. Finally, the level of experience of the project team and their use of tool support differed from one project to another, which is also likely to have affected the underlying processes.

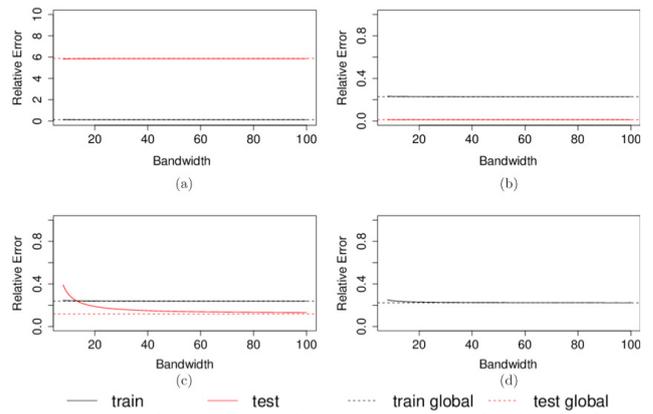

Fig. 10. Triangular models: Relative error against bandwidth for the Maxwell dataset. (a) All projects up to 1988 as the training set and 1989 as the test set. (b) All projects up to 1989 as the training set and 1990 as the test set. (c) All projects up to 1990 as the training set and 1991 as the test set. (d) Model built with all Maxwell data.

### 4.5. XBC dataset

The formation of the training and test data for the XBC dataset was modified so that it became possible to compute the relative errors for both the nonuniform and uniform models. This is because a minimum of two data points are required to compute the relative error and some specific months for XBC dataset consisted of just one project completion. While the training data was formed as described in the modeling algorithm of Sec. 3.2, the test data used in all cases for the XBC dataset consists of all projects that remained after the formation of a training set.

Figure 11 indicates a near-stationary process for all the models built using the Gaussian kernel estimator. The Epanechnikov curves are not shown here because these results are the same as those obtained from the Gaussian kernel as depicted by the similarities of their curves that attained the same relative error results for both the uniform models and the nonuniform models, respectively. Overall, the models' results according to relative error were satisfactory, as all the models resulted in a relative error value of less than 1. The predictions based on both the nonuniform models and the uniform models are almost identical for all the models developed for the XBC dataset as shown by the respective kernel type curve. These predictions are relatively good with a relative error of less than 1 for all the models.

Model accuracy (for both nonuniform and uniform models) is observed to have improved from Fig. 11(a) to Fig. 11(c) upon the introduction of new projects (data) in the building of subsequent models, thus the additional data had a positive effect on model accuracy. In Figs. 11(c)–11(f), the accuracy remained the same for all the models across time, implying that the introduction of new projects to build subsequent models had no impact on model accuracy. The predictions based on the models differ across time as was observed for all the previous datasets.

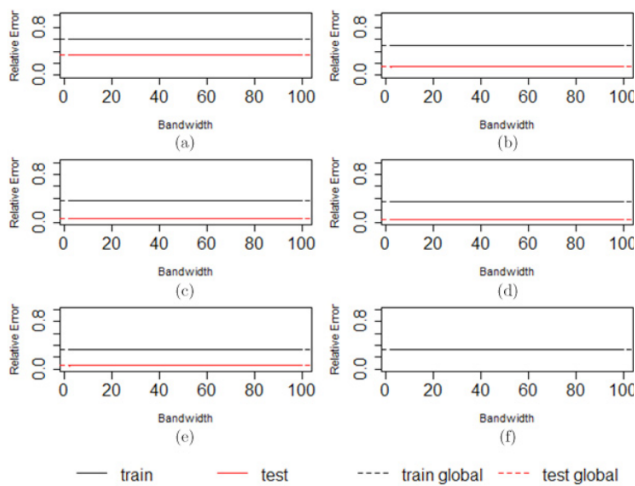
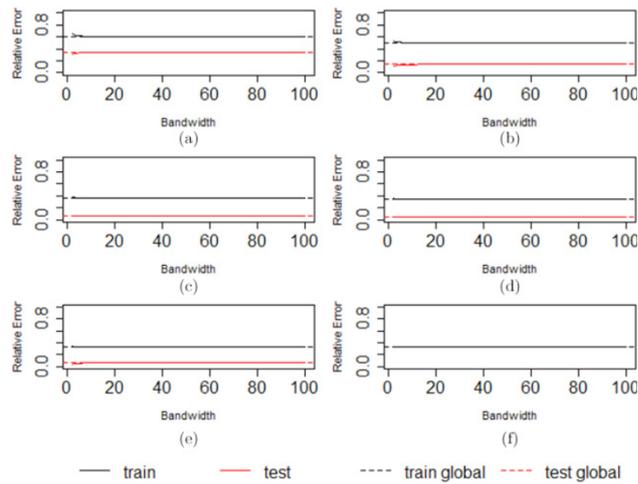

Fig. 11. Gaussian models: Relative error against bandwidth for the XBC dataset. (a) First seven projects as the training data and the rest as the test data. (b) First 10 projects as the training data and the rest as the test data. (c) First 12 projects as the training data and the rest as the test data. (d) First 13 projects as the training data and the rest as the test data. (e) First 14 projects as the training data and the rest as the test data. (f) Model built with all XBC data.

Fig. 12. Triangular models: Relative error against bandwidth for the XBC dataset. (a) First seven projects as the training data and the rest as the test data. (b) First 10 projects as the training data and the rest as the test data. (c) First 12 projects as the training data and the rest as the test data. (d) First 13 projects as the training data and the rest as the test data. (e) First 14 projects as the training data and the rest as the test data. (f) Model built with all XBC data.

Figure 12 depicts the models developed using the Triangular kernel for this dataset. Figs. 12(a) and 12(b) exhibit some levels of nonstationarity. Mapping the bandwidth of about 10 on these curves to that of Fig. 3(e) indicates that none of these models will attain stationarity over the duration of the projects that constitute the XBC dataset. The models developed for the XBC datasets, with the exception of the first two models obtained when using the Triangular kernel function, all exhibit stationarity.

The predominantly stationary processes underlying the models of the XBC dataset are largely expected as the projects were developed within a very short period by one organization, the applications were also similar and they utilized just two related programming languages. Some of the projects were also managed by the same project managers which may well have contributed to the similar characteristics found in this dataset. It is also worth noting that the presence of two models depicting nonstationarity (in the case of the Triangular kernel) suggests that there could be a subset of projects that are sufficiently different from the others to require distinct estimation functions.

## 5. DISCUSSION

Based on the results just presented, the research questions are answered as follows.

**RQ1.** Do nonuniform models (using nonuniform weighting) provide more accurate software effort estimates than uniform models (using uniform weighting) when applied over time?

In seeking to answer the research question RQ1, our results are mixed. For datasets (and therefore models) that reflected nonstationary processes, the answer is no: the uniform models are more accurate than the nonuniform models; for the datasets and models associated with stationary processes, the nonuniform and uniform models are essentially equivalent (though in the case of the Desharnais dataset, for some of the models, the nonuniform models are slightly more accurate than the uniform models). Based on the datasets analyzed in this paper, it can be concluded that, where datasets exhibit nonstationary processes, uniform models are likely to be more accurate than the nonuniform models.

However, where datasets exhibit stationary processes, the nonuniform and uniform model results are likely to be similar in terms of accuracy, and there may be some specific instances wherein the nonuniform models are better. These results are consistent across all three nonuniform kernel types.

**RQ2.** Does nonstationarity of software engineering datasets affect the accuracy of effort estimation models when applied over time?

In considering the above results, we determine that the answer to the research question RQ2 is yes. For all datasets that exhibited nonstationary processes, the models (nonuniform models) resulted in relatively large relative errors especially prior to convergence to the uniform models. In contrast, the estimation accuracy for datasets that exhibited stationarity is in almost all cases the same as that obtained from the uniform models. These results are observed for all kernel types. Thus, we would conclude that the accuracy of effort estimation models is indeed affected by the stationarity of the datasets.

**RQ3.** What is the effect of different bandwidths (if any) on software effort estimation model accuracy?

Based on the results of this study, we would assert that the effect of bandwidth on model performance is reflective of the underlying software engineering processes. It can be deduced from the results obtained for the datasets that exhibit nonstationary processes that the effect on performance is more pronounced at lower bandwidths as the associated errors are higher, while at higher bandwidths, the errors in most cases start converging to those of the uniform models. For all the datasets that exhibited stationary characteristics, the effect of bandwidth over model performance is nearly uniform across all the bandwidths.

**RQ4.** Does kernel type affect the accuracy of software effort estimation models?

For the datasets that have been analyzed in this study, the evidence indicates that the type of kernel does not affect model accuracy. The accuracies of the models as measured by the relative error were mostly the same for the respective datasets for all kernel types. The estimations based on the models using the test sets were also the same for each dataset irrespective of the kernel type that was used in the generation of the weights. This study, therefore, reaffirms the result of Kocaguneli et al.'s [29] study that did not find variation in model accuracy due the type of kernel.

In terms of using different kernel types to assess the stationarity of a dataset, there were just a few occurrences where the different kernel types generated contrasting results, as presented in Sec. 4.

# 6. THREATS TO VALIDITY

Construct validity: In meeting the requirements for construct validity of this study, four of the datasets used in this study can be considered as benchmark datasets for the software effort estimation studies for which the variables that constitute each of the datasets have undergone extensive examination in multiple prior studies. The fifth dataset, XBC, that was sourced from a multinational organization has also been used in previous studies successfully. All the preprocessing steps used are popular ones derived from the empirical software engineering domain and it is, therefore, believed that these measures reflect the practice of researchers in the software engineering domain.

External validity: The generalization of the results of this study may not hold in all situations, as the datasets used are convenience sampled from the PROMISE repository, complemented by one proprietary dataset. Though these datasets cannot be considered as representative of the entire software industry, those stored in the PROMISE repository have rather become benchmark datasets in empirical software engineering. Moreover, the five datasets were selected in terms of their possessing different characteristics. As such, these results provide promising insights into the derivation of the nature of processes underlying software engineering datasets, and the effect of stationarity on the effectiveness of nonuniform or uniform estimation models.

Internal validity: Threat to internal validity arises from our use of weighted linear regression. Considering that there are several such options that could be used to build estimation models, it is likely that the choice of modeling algorithm could have affected the results. However, this study does confirm, for the most part, the results obtained in a prior study by Kocaguneli et al. [29] in their conclusion that the choice of the kernel does not affect the accuracy of such models.

Conclusion validity: The results of this study might suffer from conclusion validity due to the diverse datasets used, however, the authors automated all the modeling processes, there was no manual tuning to affect the uniformity of the processes. One of the findings or results of this study was similar to a prior study by Kocaguneli et al. [29] thus providing evidence that our results can be trusted.

# 7. CONCLUSIONS

Five datasets including four in the public domain and one from a proprietary provider have been analyzed based on the sequential accumulation of project observations. Three kernel estimators were employed to generate nonuniform weights that were then used in regression to develop nonuniform models, where projects were given different weightings to reflect their perceived influence based on how close in time they were to a target new project. In essence, more recently completed projects were assigned higher weights in order to more strongly influence the estimation of effort for the target project. These nonuniform models were then compared with their equivalent uniform models (where no weights are applied). Graphs of relative error against bandwidth were generated, from which the nature of the curves were interpreted to determine the stationarity or otherwise of the processes underlying the datasets.

Taken overall, our results lead us to conclude that, so far as a dataset exhibits a nonstationary process, and irrespective of the kernel estimator being used, uniform models are more accurate than nonuniform models, while for a stationary process, nonuniform models are at least as good as their corresponding uniform models. The results also indicate that datasets that exhibit stationary characteristics result in better models than their nonstationary counterparts (where "better" means more accurate).

The effect of bandwidth was found to be associated with the underlying process used in generating the datasets. There is no difference in relative error when the datasets exhibit stationarity across the different bandwidths. In contrast, for a nonstationary process, lower bandwidths may result in larger relative errors while at higher bandwidths, relative errors tend to decrease until they converge to those of the uniform models.

The evidence drawn from this study further suggests that the accuracy of models is independent of the kernel type used in the generation of weights for the nonuniform models. This is observed in the fact that, for each dataset, all three kernel estimators resulted in the same relative errors for all equivalent models and their estimates for the test set observations.

In looking ahead, the use of the kernel estimator functions to develop effort estimation models could in fact contribute more generally to the practice of software engineering as it would be possible to identify changes in the underlying processes that might not be apparent to project managers. This could inform the provision of appropriate remedies where these underlying changes are not favorable to the software engineering process. In order to use the kernel estimator to provide good effort estimates as part of a proactive management process, the choice of the kernel bandwidth is critical, and this will require the evaluation of various bandwidth selection methods in a software effort estimation context. To the best of our knowledge, none of these "optimum" bandwidth selection methods have been applied to software engineering effort data and as such, their efficacy is at present not known in this regard. In terms of this study this was not of concern, as (i) the objective of this study was the identification of stationarity in software engineering datasets; and (ii) the analysis was conducted retrospectively on "complete" datasets. However, this comparison should be undertaken as a future research activity, particularly if the method is to be applied to "live" growing datasets.